\documentclass[12pt]{article} \topmargin= 0.3cm \textwidth 16.5cm
=0cm \headsep =0cm

\usepackage {graphicx}

\begin{document}

\title{Gravity-induced instability and gauge field localization}
\author{K. Farakos \footnote{kfarakos@central.ntua.gr} and P. Pasipoularides  \footnote{paul@central.ntua.gr} \\
       Department of Physics, National Technical University of
       Athens \\ Zografou Campus, 157 80 Athens, Greece}
\date{ }
       \maketitle

\begin{abstract}
The spectrum of a massless bulk scalar field $\Phi$, with a possible
interaction term of the form ${\cal L}_{int}=-\xi R\; \Phi^{2}$, is
investigated in the case of RS-geometry \cite{Ran}. We show that the
zero mode for $\xi=0$, turns into a tachyon mode, in the case of a
nonzero negative value of $\xi$ ($\xi<0$). As we see, the existence
of the tachyon mode destabilizes the $\Phi=0$ vacuum, against a new
stable vacuum with nonzero $\Phi$ near the brane, and zero in the
bulk. By using this result, we can construct a simple model for the
gauge field localization, according to the philosophy of Dvali and
Shifman (Higgs phase on the brane, confinement in the bulk).
\end{abstract}

\section{Introduction}
It is well known that theories which are believed as fundamental,
i.e. the string theory, are formulated in multidimensional spaces.
For this reason there has been a great interest in field theory
models with more than three spatial dimensions. The standard way to
retain four-dimensional physics in multidimensional-models is to
assume that the extra dimensions are compactified to a large scale
$M$. However, in resent years, there is an alternative scenario
which has attracted attention. This scenario is based on the idea
according to which we are trapped in a submanifold with three
spatial dimensions (brane world) that is embedded in a fundamental
multi-dimensional manifold (bulk). A new feature of this scenario,
is that it allows the extra dimensions to be large or even infinite.

In this paper we will concentrate our attention on the interesting
case of the RS2-brane world scenario of Ref. \cite{Ran}. In this
variation of the brane world scenario, we have a single brane with a
positive energy density (the tension $\sigma$) and a negative,
fine-tuned, five dimensional cosmological constant $\Lambda$. This
model implies a non-factorizable space time geometry of the form of
$AdS_{5}$ around the brane (see Eq. (5) below). The four-dimensional
particles, in this model, are expected to be gravitationally trapped
on the brane. We see in Refs. \cite{Ran,Bajc} that gravitons and
scalar fields exhibit a normalizable zero mode plus a continuous
spectrum\footnote{In this case there is no mass gap. However, the
continuous modes interact very weakly with the four dimensional
matter (zero mode). In this way the continuous modes do not affect
the low energy four dimensional physics}. Unfortunately there is no
normalizable zero mode in the case of fermions and gauge
fields\footnote{There is a normalizable bound state for fermions
only if the brane tension in negative (or $\sigma<0$). Gauge fields
are localized neither on a brane with positive tension nor on a
brane with negative tension.}, for the RS2-model, see Ref.
\cite{Bajc}.

According to the above discussion, for achieving localization of the
fermion or gauge field we must invoke an alternative scenario or an
alternative mechanism. There are several variation of the brane
world scenario, with different geometries than that of the RS2-model
(different warp factors) or with more than five dimensions, or even
in flat space time with topological defects, where the fermion or
gauge field localization has been achieved, see Refs.
\cite{Rub,Keh,Ner:Giov} and references there in. However, as it is
emphasized in Ref. \cite{Rub,Rumm}, the most difficult task is the
massless non-Abelian gauge field localization on the brane, as any
acceptable mechanism for this must preserve the charge universality.

A non gravitational mechanism, which seems to solve this problem in
a clever way, is the Dvali-Shifman mechanism of Ref. \cite{Shif}.
This mechanism is based on the idea of the construction of a gauge
field model which exhibits a non-confinement phase on the brane and
a confinement phase on the bulk. Thus gauge fields, and more
generally fermions and bosons with gauge charge (see Ref.
\cite{Rub}), can not escape into the bulk unless we give them energy
greater than the mass gap $\Lambda_{G}$, which emerges from the
nunpurturbative confining dynamics of the gauge field model in the
bulk. For criticism or a better insight on this mechanism see Refs.
\cite{Rub,Rumm,Ark,Tetr}. Also, the Dvali-Shifman mechanism in the
case of conformal matter has been studied in Ref. \cite{Dem}.

From the point of view of lattice, it is worth to note that there
are some models with anisotropic couplings, in multi-dimensional
spaces, that exhibit a new phase, the layer phase as it is named.
The main feature of the layer phase is a non-confinement phase along
the layer combined with a confinement along the extra dimensions.
The idea of the layer phase was originated by Fu and Nielsen, in
Ref. \cite{fu}. Recent research aiming at the investigation of the
layer phase for several models with anisotropic coupling, as a way
to achieve non-Abelian gauge field localization, can be found in
Refs. \cite{lat6:KA,lat1,lat3,lat4}. Particularly, in Ref.
\cite{lat3} the anisotropic SU(2) adjoint Higgs model is analyzed.
In addition, the U(1) model in the RS-background is analyzed in Ref.
\cite{lat1}. We see that the lattice results indicate a layer phase
for both of these models.

In this paper we aim at the construction of a model where the
Dvali-Shifman mechanism is triggered by the geometry of the
multidimensional space time, and not by an auxiliary neutral scalar
field which forms a kink topological defect toward the extra
dimension, as is done in the original paper of Ref. \cite{Shif}. For
this we have considered an SU(2) gauge field model with a scalar
triplet, in the RS-geometry. In this model we have included a
possible interaction term of the form ${\cal L}_{int}=-\xi R
\Phi^{2}$, see Refs. \cite{R1,R2,Haw}, where $\xi$ is a
dimensionless numerical factor. In section 4 we study how the
spectrum of the scalar field is modified when this additional
interaction turns on. We see that the zero mode, for $\xi=0$, turns
into a tachyon mode (a normalizable bound state with negative
energy), in the case of a nonzero negative value of $\xi$ ($\xi<0$).
The tachyon mode implies an instability of the $\Phi=0$ vacuum
against a new stable vacuum for the scalar field, which is described
by a $\Phi(z)$ configuration with nonvanishing value for $\Phi$ near
the brane ($z$ is the extra dimension). Thus, in the background of
this stable configuration for the scalar field, the gauge symmetry
is valid in the bulk (confinement phase), but it is spontaneously
broken on the brane (non-confinement phase).

It is worth to note that in Refs. \cite{Haw,ford} the same non
minimally coupled model is considered, for four dimensions in the
case of de Sitter space-time. It is obtained that for a specific
range of values of $\xi$ the scalar field is rendered unstable, and
this result have straightforward implications to the cosmological
constant problem. The philosophy of our paper is quite closely to
that of the previous works of Refs. \cite{Haw,ford}. However there
are some important differences, as in the present work we study the
five dimensional space-time in the present of the RS-metric, and we
use our results in order to argue for the localization of the gauge
fields on the brane.

\section{Gravity-induced instability for the scalar field vacuum $\Phi=0$}

We consider an SU(2) gauge field model with a scalar triplet
$\Phi^{c}$ ($c=1,2,3$), in the adjoint representation, in the
presence of gravity. The action of this model is:
\begin{equation}
 S=\int d^{5}x \;({\cal L}_{gravity}+{\cal L}_{gauge}+{\cal L}_{scalar})
\end{equation}
The volume element is $d^{5}x=d^{4}x dx_{4}$, where $x_{4}$ is the
infinite extra dimension. In what will follow we set $x_{4}=z$.

The Lagrangian for gravity is chosen to be identical with that of
the Randall-Sundrum model \cite{Ran}, which is given by the equation
\begin{equation}
{\cal L}_{gravity}=-\frac{1}{16 \pi
G_{5}}\sqrt{|g|}R+\Lambda\sqrt{|g|}+\sigma\delta(z)\sqrt{|g^{(brane)}|}
\end{equation}
where $\sigma$ is the brane tension, $\Lambda$ is the
five-dimensional cosmological constant, and $G_{5}$ is the
five-dimensional Newton 's constant. In addition $R$ is the five
dimensional Ricci scalar, $g$ is the determinant of the five
dimensional metric tensor $g_{MN}$ ($M, N=0,1,...,4$), and
$g^{brane}$ is the determinant of the metric tensor on the brane,
which is defined by the equation
$g^{(brane)}_{\mu\nu}=g_{\mu\nu}(z=0)$ ($\mu,\nu=0,1,...,3$). For
the signature of the metric $g_{MN}$ we adopt the convention
$(+,-,-,-,-)$, and for the sign of the curvature tensor we adopt the
convention of Ref. \cite{R1}
($R^{\alpha}_{\beta\gamma\delta}=\partial_{\delta}\Gamma^{\alpha}_{\beta\gamma}-...$).

The Lagrangians for the gauge field and the scalar field, in curved
space-time, are
\begin{eqnarray}
{\cal L}_{gauge}&=&-\frac{1}{4}\sqrt{|g|}F^{c}_{M N}F^{c}_{K L}\;g^{M K}g^{N L}\\
{\cal L}_{scalar}&=&\sqrt{|g|}\left(\frac{1}{2}g^{M N} D_{M}\Phi_{c}
D_{N}\Phi_{c}-V(\Phi)\right)
\end{eqnarray}
In what follows we assume that $V(\Phi)=\lambda
(\Phi\cdot\Phi)^{2}$, or we assume that there is not a mass term.
Also we have used the notation $\Phi=(\Phi_{1},\Phi_{2},\Phi_{3})$
and $\Phi\cdot\Phi=\Phi_{1}^{2}+\Phi_{2}^{2}+\Phi_{3}^{2}$. In
addition the covariant derivative in the adjoint representation is
$D_{M}=\partial_{M}-ig_{5} A_{M}$ where $A_{M}=A_{M}^{b}T^{b}$,
$T^{b}=-i\epsilon^{bcd}$, and $g_{5}$ is the five-dimensional gauge
coupling.

If we chose $\Lambda=\frac{-4\pi}{3}G_{5}\sigma^{2}$, the
corresponding equations of motion of the Lagrangian of Eq. (1) have
a \textit{classical stable solution} of the form
\begin{equation}
ds^{2}=a^{2}(z)(dx_{0}^{2}-dx_{1}^{2}-dx_{2}^{2}-dx_{3}^{2})-dz^{2},
\quad \Phi_{c}=0, \quad A_{M}=0
\end{equation}
where $a(z)=e^{-k|z|}$ is the warp factor, with
$k=\frac{4\pi}{3}G_{5}\sigma$.

We aim to investigate the effects of a possible additional
interaction term between the scalar field and the gravity, of the
form $-\xi R\; \Phi^{2}$ (see Refs. \cite{R1,R2}), where $\xi$ is a
dimensionless numerical factor. Then ${\cal L}_{scalar}$ is modified
as
\begin{equation}
{\cal L}_{scalar}=\sqrt{|g|}\left(\frac{1}{2}g^{MN} D_{M}\Phi_{c}
D_{N}\Phi_{c}-\frac{1}{2}\xi R\ (\Phi\cdot\Phi)-V(\Phi)\right)
\end{equation}
Note that Eq. (5) remains a solution of the equations of motion even
for $\xi\neq 0$.

The spectrum of the massless scalar field, around the stable
solution of Eq. (5) for $\xi=0$, consists of a zero-mode plus a
continuous tower of states (see for example Ref. \cite{Bajc}). In
section 4 we will show analytically that the zero mode for $\xi=0$,
turns into tachyon mode (a normalizable bound state with negative
energy), in the case of a nonzero negative value of $\xi$ ($\xi<0$).

The existence of a localized tachyon mode around $z=0$, when we turn
on the interaction term $-\xi R \Phi^{2}$, renders the solution of
Eq. (5) unstable. Due to the instability, the vacuum expectation
value of the scalar field can not be zero everywhere anymore. The
scalar field vacuum will be a function of $z$, which form is
expected to be similar with that of the tachyon mode, see Eq. (32)
in section 4. This suggests a nonzero value for $\Phi_{3}(z)$ near
the brane, which tends rapidly to zero, outside in the bulk. Note,
that the nonzero value of the scalar field near the brane is
stabilized by the interaction term $\lambda (\Phi\cdot\Phi)^{2}$.

Now will try to give a brief description of the new stable metric of
the nonminimally coupled theory with negative $\xi$. In this case
the new stable metric is expected to be of the form of Eq. (5) (or
Eq. (10)). However, the behavior of the warp factor is not given by
the equation  $a(z)=e^{-k|z|}$ anymore. The exact form for $a(z)$
can be found only if we solve the system of the equations of motion
for the lagrangian of the system (see Eqs. (6) and (2)) with two
unknown functions $a(z)$ and $\Phi(z)$ and the appropriate boundary
conditions. Note that as $\Phi(z)\rightarrow 0$ for $|z|\rightarrow
+\infty$, the warp factor for $|z|\rightarrow +\infty$ must be of
the form $a(z)=e^{-k|z|}$ or $a(z)=e^{k|z|}$, as these two functions
satisfy the equations of motion if we set $\Phi(z)=0$. Otherwise in
this paper we have not determined analytically or numerically the
metric of the new stable solution, and we have left this problem for
future investigation.

\begin{figure}[h]
\begin{center}
\includegraphics[scale=1.5,angle=0]{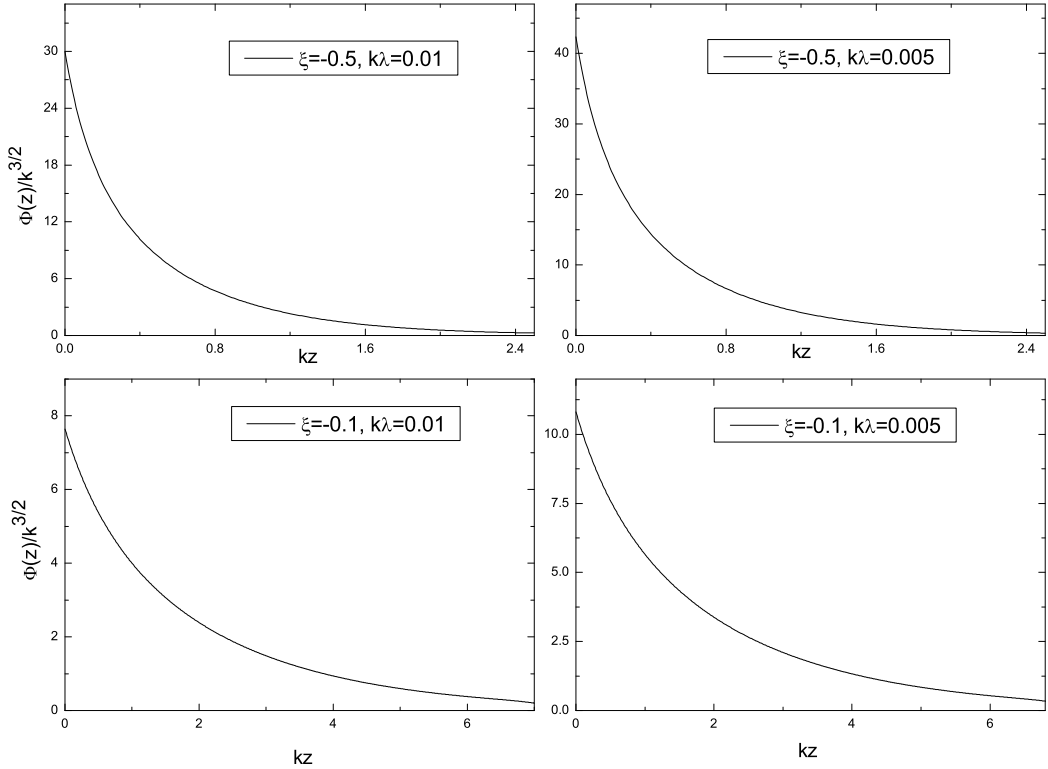}
\end{center}
\caption {The scalar field vacuum $\Phi(z)/k^{3/2}$ versus $z$
($z>0$ as $\Phi(z)=\Phi(-z)$), for $k \lambda=0.01,0.005$, and
$\xi=-0.5,-0.1$, in the background of the RS-metric.} \label{1}
\end{figure}

However, in order to estimate quantitatively the scalar field vacuum
as a function of z we will assume that the RS-metric is fixed "by
hand". We have solved numerically the equation of motion which
corresponds to the Lagrangian of Eq. (6) for $\xi<0$, in the
background of the RS-metric, and we have plotted our results in Fig.
\ref{1} for $z\geq 0$, as $\Phi(z)=\Phi(-z)$. We see that indeed the
scalar field vacuum is nonzero on the brane and zero in the bulk, as
expected from the behavior of Ricci scalar R versus $z$ (or $w$)
(for details see the next section). In particular, we solved
numerically the ordinary differential equation:
\begin{eqnarray}
-\Phi''(z)+4 k \Phi'(z)+20 k^{2}|\xi|\Phi(z)+4 \lambda \Phi^{3}(z)=0
\end{eqnarray}
for $z>0$, with the boundary conditions
\begin{eqnarray}
\partial_{z}\Phi(0)+8 k |\xi| \Phi(0)=0, \quad \Phi(+\infty)=0
\end{eqnarray}
The first boundary condition comes from the $\delta(z)$ term in
Riccy scalar R of Eq. (12) below. In addition, we emphasize that the
potential term $\lambda (\Phi\cdot\Phi)^{2}$ is necessary in order
to achieve a behavior like that of Fig. \ref{1}.

In the framework of the above discussion we can argue in favor of a
possible scenario for the gauge field localization on the RS2-brane
world. The nonzero value \footnote{We assume that the scalar field
is directed toward the c=3 direction in the isospin space.} of
$\Phi_{3}$ field, for $z\approx 0$, implies a non-confinement phase
on the brane (or the SU(2) symmetry is spontaneously broken to U(1)
on the brane). On the other hand, the zero value of $\Phi_{3}$
outside the brane implies a confinement phase into the bulk, (or the
SU(2) symmetry is restored in the bulk). In this way the gauge field
of U(1) theory (photon) is localized on the brane, as for escaping
in the bulk, it requires energy equal to $\Lambda_{gap}$, where
$\Lambda_{gap}$ is the mass gap emerging from the nonperturbative
confining dynamics of the SU(2) gauge field theory, in the bulk.
Note, that in principle, the same mechanism can be used for a model
with an arbitrary gauge group $G$ in the bulk which breaks to a
subgroup $G'$ on the brane. Similarly, the gauge fields related to
$G'$ are localized on the brane and are separated by a mass gap
$\Lambda_{gap}$ from the bulk modes, which are massive because of
confinement (also see Refs. \cite{Rumm,lat3}).

\begin{figure}[h]
\begin{center}
\includegraphics[scale=1.2,angle=0]{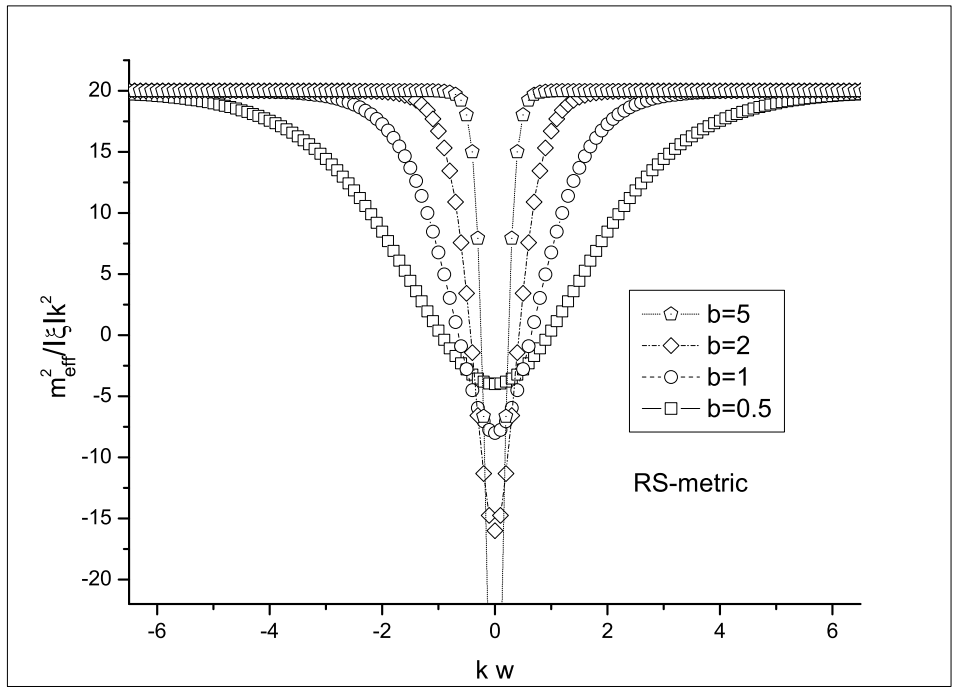}
\end{center}
\caption {The effective mass $m_{eff}^{2}=-|\xi|R$ versus the extra
dimension for the RS-metric of Eq. (8). In order to plot the delta
function we used the representation $\delta(w)=\lim_{b\rightarrow
+\infty}b\; sech(bw)^{2}$.} \label{2}
\end{figure}

\section{Ricci scalar as an effective z(or w)-dependent
mass term for the scalar field}

In this section we present an alternative way to establish
qualitatively the picture of the nonvanishing $\Phi$-condensation
near the brane, and the $\Phi=0$ in the bulk.

It is convenient to make the change of variable
\begin{equation}
w=sgn(z)\frac{(e^{k|z|}-1)}{k}
\end{equation}
Then the metric of Eq. (5) can be put into the manifestly conformal,
to the five-dimensional Minkowski space, form
\begin{equation}
ds^{2}=\alpha^{2}(w)(dx_{0}^{2}-dx_{1}^{2}-dx_{2}^{2}-dx_{3}^{2}-dw^{2})
\end{equation}
where $\alpha(w)=1/(k|w|+1)$

We can show that for the above metric of Eq. (10) the Ricci scalar
is
\begin{equation}
R(w)=-4\left(2\frac{\alpha''(w)}{a^{3}(w)}+\frac{(\alpha'(w))^{2}}{\alpha^{4}(w)}\right)
\end{equation}
and for  $\alpha(w)=1/(k|w|+1)$ we obtain that
\begin{equation}
R(w)=16 k \delta(w)-20 k^{2}
\end{equation}
\textit{Note the opposite signs of the delta function and the
constant term in the above equation.}

\begin{figure}[h]
\begin{center}
\includegraphics[scale=1.2,angle=0]{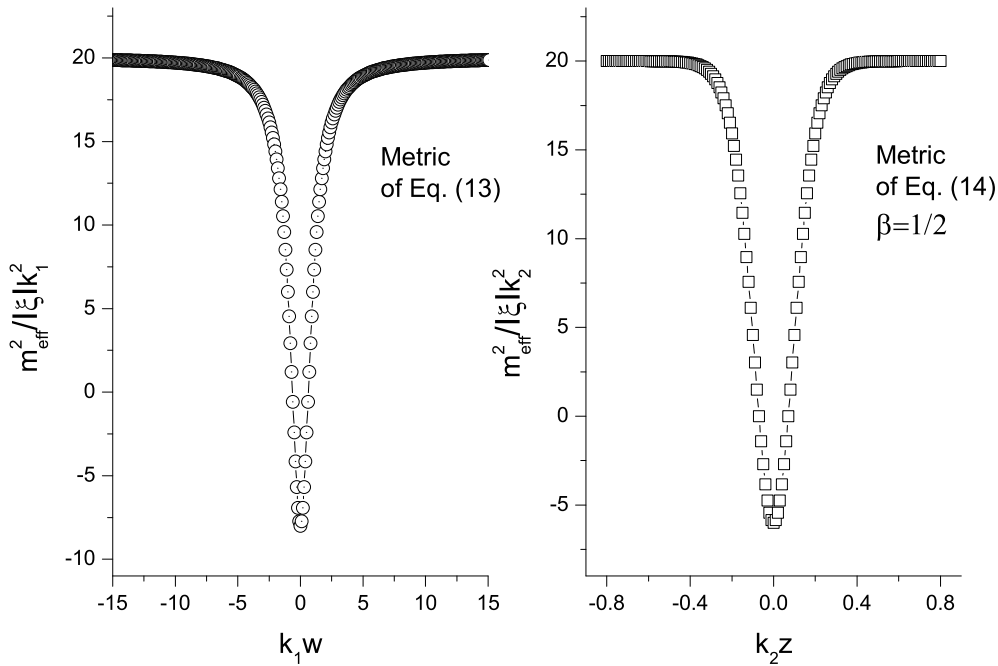}
\end{center}
\caption {The effective mass $m_{eff}^{2}=-|\xi|R$ versus the extra
dimension for the metrics of Eqs. (13) and (14).} \label{3}
\end{figure}

From the lagrangian of Eq. (6) we see that the interaction term
$-\xi R \Phi^{2}$, can be viewed as an effective w-dependent mass
term, which corresponds to the scalar field $\Phi$ (for more details
see Ref. \cite{Gu} and references there in). We see that, for
$\xi<0$, outside the brane the effective mass
$m_{eff}^{2}(w)=-|\xi|R$ is positive (see Eq. (12)), and thus the
SU(2) symmetry is not violated. On the other hand, for $w\approx 0$,
the existence of the delta function renders the value of
$m_{eff}^{2}(w)$ negative on the brane (particularly
$m_{eff}^{2}(w)=-\infty$ on the brane), and thus the SU(2) symmetry
is spontaneously broken.

As a figure with a delta function may cause same confusion, we will
use the representation $\delta(w)=\lim_{b\rightarrow +\infty}b\;
sech(bw)^{2}$, in order to plot the delta function. In Fig. \ref{2}
we have plotted the effective mass $m_{eff}^{2}(w)/|\xi| k^{2}$
versus $k w$, for several values of $b$. In this figure we see that
$m_{eff}^{2}(w)$ is negative near the brane and positive in the
bulk, as is expected.

It is worth to note that the mechanism of this paper is expected to
work also for metrics other than that of Eq. (5). For this we have
assumed the two metrics with smooth warp factors, of Eqs. (13) and
(14). We see that for large values of the extra dimension both of
them tend asymptotically to an $AdS_{5}$ spacetime. The metric of
Eq. (13) has been found in Ref. \cite{Giov1} by M. Giovannini in the
case of an additional Gauss-Bonnet term in the gravity action. The
metric of Eq. (14) has been found in Ref. \cite{Keh} by A. Kehagias
and K. Tamvakis, in the case of a bounce type solution in the extra
dimension  (see their paper for a description of the parameters in
Eq. (13)). Fig. \ref{2} confirms that indeed the effective mass, for
these two metrics, is negative on the brane and positive in the
bulk, as is needed for the above presented mechanism to work.
\begin{eqnarray}
ds^{2}&=&\alpha^{2}(w)(dx_{0}^{2}-dx_{1}^{2}-dx_{2}^{2}-dx_{3}^{2}-dw^{2}), \quad \alpha(w)=\frac{1}{\sqrt{k_{1}^{2}w^{2}+1}} \\
ds^{2}&=&a^{2}(z)(dx_{0}^{2}-dx_{1}^{2}-dx_{2}^{2}-dx_{3}^{2})-dz^{2},
\quad a(z)=\frac{e^{-\beta
\tanh^{2}(k_{2}z)}}{(\cosh^{2}(k_{2}z))^{2\beta}}
\end{eqnarray}

\section{The tachyon mode for the scalar field}

In this section we study in detail the spectrum of the scalar field
in the case of the additional interaction term  $-\xi R \;\Phi^{2}$.
\begin{figure}[h]
\begin{center}
\includegraphics[scale=1.2,angle=0]{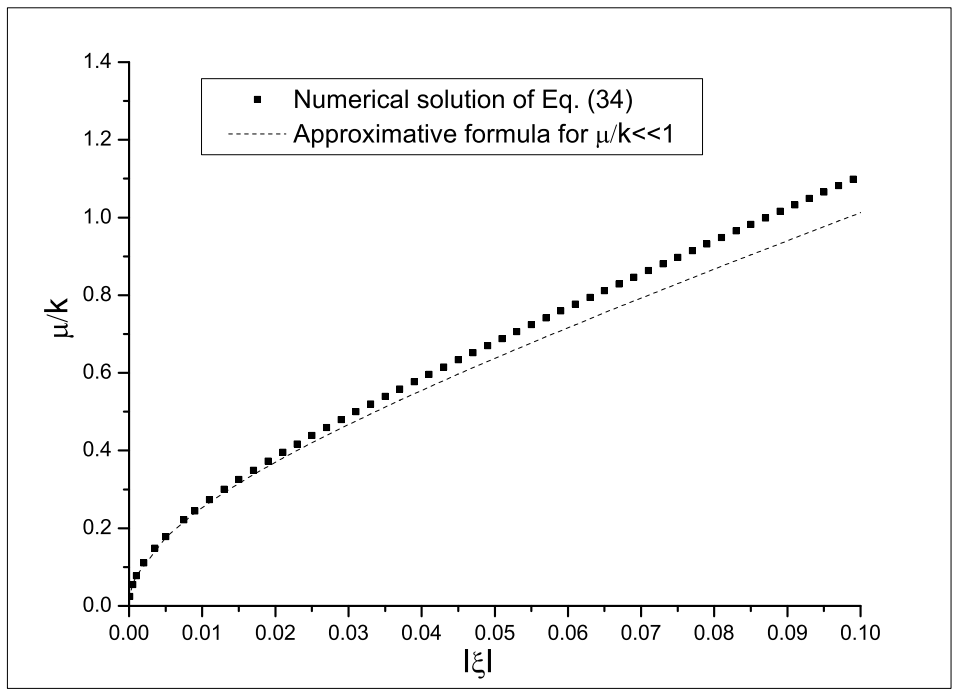}
\end{center}
\caption {$\mu/k$ versus k. The continuous line corresponds to the
approximative formula of Eq. (35) for $\mu/k<<1$, and the discrete
points to the exact solution of Eq. (34), as it is found
numerically. We see that indeed the formula of Eq. (35) gives the
correct result for $|\xi|<<1$.} \label{4}
\end{figure}
We consider a small perturbation $\Phi_{3}$, directed toward the
$c=3$ direction in the isospin space, around the solution of Eq.
(5). The corresponding linearized equation of motion is
\begin{equation}
\frac{1}{\sqrt{|g|}}\;
\partial_{M}\left[\sqrt{|g|}g^{MN}\partial_{N}\Phi(x,w)\right]+\xi
R(w) \;\Phi(x,w)=0
\end{equation}
where, for the sake of simplicity, we have dropped the isospin index
from the scalar field. In addition
$g^{MN}=(1/\alpha^{2}(w))(\eta^{\mu\nu},-1)$ ($\mu,\nu=0,1,2,3$)
where $\eta^{\mu\nu}=(1,-1,-1,-1)$, and $\sqrt{|g|}=\alpha^{5}(w)$.
Eq. (15) can be written as
\begin{equation}
\alpha^{3}\eta^{\mu\nu}\partial_{\mu}\partial_{\nu}\Phi-\partial_{w}(\alpha^{3}\partial_{w}\Phi)+\alpha^{5}\xi
R \;\Phi=0
\end{equation}
We are looking for solutions of the form
\begin{equation}
\Phi(x,w)=\phi(x)\frac{\psi(w)}{\alpha^{3/2}(w)}
\end{equation}
where
\begin{equation}
\eta^{\mu\nu}\partial_{\mu}\partial_{\nu}\phi(x)=-m^{2}\phi(x)
\end{equation}
or $\phi(x)\sim e^{ipx}$, and $m^{2}=p_{\mu}p^{\mu}$ is the four
dimensional mass.

The function $\psi(w)$ verifies the Schrondiger like equation
\begin{equation}
-\psi''(w)+\left[V(w)-m^{2}\right]\psi(w)=0
\end{equation}
where the potential $V(w)$ is defined as
\begin{equation}
V(w)=\frac{(\alpha^{3/2}(w))''}{\alpha^{3/2}(w)}+\xi\alpha^{2}(w)R(w)
\end{equation}
Note that Eq. (19) determines the four dimensional particle spectrum
of the theory.

For the metric of Eq. (10) we obtain
\begin{equation}
\frac{(\alpha^{3/2}(w))''}{\alpha^{3/2}(w)}=-3 k
\delta(w)+\frac{15}{4} k^{2}\frac{1}{(k|w|+1)^{2}}
\end{equation}
Thus from Eqs. (12), (20) and (21) we obtain
 \begin{equation}
V(w)=-k(3-16\xi)\delta(w)+k^{2}(\frac{15}{4}-20\xi)\frac{1}{(k|w|+1)^{2}}
\end{equation}
If we set $y=kw$ we obtain
\begin{equation}
-\psi''(y)+\left[\overline{V}(y)-\frac{m^{2}}{k^{2}}\right]\psi(y)=0
\end{equation}
The boundary conditions on the brane are
\begin{eqnarray}
\psi(+0)-\psi(-0)&=&0\\
\partial_{y}\psi(+0)-\partial_{y}\psi(-0)&=&-b \psi(0)
\end{eqnarray}
where the second condition arises due to the existence of the delta
function in Eq. (22) (or Eq. (29)). For the definition of b see Eq.
(27) below. The potential $\overline{V}(y)$ is given by Eq. (29)
below.

As we have emphasized in previous sections we  expect a tachyon mode
for $\xi<0$ (or $|\xi|=-\xi$). If we set $m^{2}=-\mu^{2}<0$ we
obtain.
\begin{equation}
-\psi''(y)+\left[\overline{V}(y)+\frac{\mu^{2}}{k^{2}}\right]\psi(y)=0
\end{equation}
In addition, if we set
\begin{eqnarray}
b&=&(3+16|\xi|) \\ c&=&(\frac{15}{4}+20|\xi|)
\end{eqnarray}
the potential $V(y)$ can be written in the form
\begin{equation}
\overline{V}(y)=-b\delta(y)+\frac{c}{(|y|+1)^{2}}
\end{equation}
If we search for a normalizable solution of Eq. (26) we obtain that
\begin{eqnarray}
\psi(y)&=&N\sqrt{y+1}\;K_{\nu}\left[\frac{\mu}{k}(y+1)\right] \quad if \quad y\geq 0 \\
\psi(y)&=&N\sqrt{-y+1}\;K_{\nu}\left[\frac{\mu}{k}(-y+1)\right]
\quad if \quad y\leq 0
\end{eqnarray}
where $K_{\nu}(x)$ are the modified Hankel functions, and
$\nu=\sqrt{c+\frac{1}{4}}=\sqrt{4+20|\xi|}$.

Note that if take into account Eq. (9), then
\begin{eqnarray}
\widetilde{\psi}(z)=N
e^{\frac{k|z|}{2}}\;K_{\nu}\left[\frac{\mu}{k}e^{k|z|}\right]
\end{eqnarray}
From the boundary condition (25), if we take into account Eqs. (30)
and (31) we obtain
\begin{equation}
(b+1)K_{\nu}(\frac{\mu}{k})+2 \frac{\mu}{k}K'_{\nu}(\frac{\mu}{k})=0
\end{equation}
If we use the identity
$K'_{\nu}(x)=-K_{\nu-1}(x)-\frac{\nu}{x}K_{\nu}(x)$ we obtain
\begin{equation}
(b+1-2
\nu)\frac{K_{\nu}(\frac{\mu}{k})}{(\frac{\mu}{k})K_{\nu-1}(\frac{\mu}{k})}=2
\end{equation}
For $\mu<<k$, if we take into account the asymptotic formula
$K_{\nu}(x)\approx \frac{1}{2}
\Gamma(\nu)\left(\frac{2}{x}\right)^{\nu}$ as $x\rightarrow 0$, we
find the solution
\begin{equation}
\mu^{2}=4 k^{2} (1+4 |\xi|-\sqrt{1+5|\xi|})(2 \sqrt{1+5 |\xi|}-1)
\end{equation}
and for $|\xi|<<1$ we find the simple expression
\begin{equation}
\mu^{2}\approx k^{2} (6|\xi|+\frac{85}{2} \xi^{2})
\end{equation}
Note that for $\xi=0$ we obtain $\mu=0$, as is expected.

In Fig. \ref{4} we compare the approximative formula of Eq. (35)
with the exact solution of Eq. (34), as it is found by a numerical
method. We see that indeed Eq. (35) gives the correct result for
$|\xi|<<1$ (or for $\mu/k<<1$). In addition, we have checked
numerically that Eq. (34) has a unique solution for all the range of
values of $\xi<0$, which is a monotonic increasing function of
$|\xi|$. However, if $|\xi|$ is not small ($|\xi|>>0.04$ and
$\mu/k>>0.6$) as we see in Fig. \ref{3}) Eq. (35) fails to give the
correct value for $\mu/k$. For $\xi>0$ there is no tachyon mode of
the form of Eq. (32). Finally, we note that the case of $\xi>0$ may
have same interest, however it is not investigated in this paper.

\section{Conclusions}

We investigated the spectrum of a scalar field, with a possible
interaction term of the form $-\xi R \Phi^{2}$, in the background of
the RS-metric. We show that the zero mode for $\xi=0$ turns into a
tachyon mode, in the case of a nonzero negative value of $\xi$
($\xi<0$). As we argue in sections 2,3 and 4, we expect that the
tachyon mode renders the classical solution of Eq. (5) (RS-metric
plus $\Phi=0$) unstable, against a new stable solution with nonzero
$\Phi$-field condensation near the brane, and a new metric of the
form of Eq. (5) with a different warp factor (for details see
section 2).

In the framework of the above discussion we can construct a simple
model where the Dvali-Shifman mechanism is triggered by the geometry
of the multidimensional space time. The advantage of this model is
that the neutral scalar field in Ref. \cite{Shif}, which forms a
kink topological defect towards the extra dimension, is not
necessary, as it has been replaced by the RS-metric.

\section{Acknowledgements}
We are grateful to Professors A. Kehagias, G. Koutsoumbas and N.
Tetradis for reading and commenting on the manuscript. K.F thanks
Professor M. Giovannini for useful and elaborating discussions
during his visit at CERN. P.P. thanks Dr. P. Manouselis for
important discussions. This work was partially supported by "Thales"
project of NTUA and the "Pythagoras" project of the Greek Ministry
of Education-European community (EPEAEK-EKT, 25/75).

\end{document}